# Operation of the intensity monitors in beam transport lines at Fermilab during Run II[*]

**J. Crisp**[a†], **B. Fellenz**[a], **J. Fitzgerald**[a], **D. Heikkinen**[a], **M.A. Ibrahim**[a‡]

[a]*Fermi National Accelerator Laboratory,*
*Batavia, IL, U.S.A.*

*E-mail:* `cadornaa@fnal.gov`

ABSTRACT: The intensity of charged particle beams at Fermilab must be kept within pre-determined safety and operational envelopes in part by assuring all beam within a few percent has been transported from any source to destination. Beam instensity monitors with toroidial pickups provide such beam intensity measurements in the transport lines between accelerators at FNAL. During Run II, much effort was made to continually improve the resolution and accuracy of the system.

[*] Work supported by Fermi Research Alliance, LLC under Contract No. DE-AC02-07CH11359 with the United States Department of Energy.
[†] Now at Michigan State University.
[‡] Corresponding author.

# Contents

## 1. Overview

Proton beams for Tevatron operation at Fermilab originate in a Cockcroft-Walton accelerator and traverse three other accelerators before ending up at the Tevatron. Antiprotons pass through even more accelerators. Between accelerators are transfer lines to take the beam from one ring to another. To accomplish a transfer, the beam must be kicked out of the source accelerator and kicked into the destination accelerator. If the beam is poorly kicked or the beam parameters are not properly matched, part of the beam may be lost. Knowing this transfer efficiency is critical to obtaining the highest beam intensities and preventing unwanted beam losses from causing excessive radioactivation of the beam line components.

Transfer line beam intensities are measured from non-destructive, magnetically-coupled toroidal pickups. Most transport lines have been installed with a minimum of two of these intensity monitors. Each intensity monitor supplies the total beam pulse intensity measurement to the control system. These measurements allow operators to maximize the intensity of the final beam delivered to the Tevatron. The demand for higher luminosity has driven various improvements such as improved signal to noise and better repeatability.

## 2. Transformer theory review

Most transport line beam intensity monitors rely primarily on the interaction of the beam's magnetic field component with a current transformer (CT). This magnetic coupling between the beam and the CT can be described with basic transformer circuit theory [1]-[4].

As a transformer, the CT physically cannot pass dc. Whenever the output voltage is non-zero and the current is constant, as during the flat top of the beam pulse, the voltage decays toward zero exponentially. This downward slope of the top of the output voltage pulse resulting from a flattop current input pulse is described by the CT's droop rate which is related to the low frequency cut-off point.

The upper bound of the CT's frequency bandwidth is mathematically related to ratio of the core's inductance and the load resistance. Its inductance is based on several properties including the geometry of the magnetic core, the amount of air gap in the magnetic circuit, the properties of the core material (especially permeability and hysteresis), the operating temperature of the core, and whether the core is laminated to reduce eddy currents.

For frequencies at least one decade from the low and high frequency cut-off points, the beam serves as a single turn primary winding, inducing magnetic flux into the core's annular cross-section. Assuming no losses, an N-turn secondary sense winding on the core intercepts the flux in the core and generates a voltage signal across a load resistor.

## 3. System installations

The CTs used are purchased from Pearson Electronics [5]. The proper model is chosen based on the requirements demanded by the beam pipe diameter and the expected beam structure (i.e. signal bandwidth and sensitivity). Its sensitivity is driven by N-turn secondary windings and the load resistor. Within its specified frequency range, the phase shift is usually less than 6 degrees, and the amplitude error is less than 1%.

The installation of a complete intensity monitor consists of 2 main parts. One is located within the beam enclosure and includes a toroidal CT, ceramic break, gap network, single-turn calibration loop, cables, and mounting stand. The second part is located outside the enclosure within a service building or experimental hall. This includes the electronics needed for signal processing and read back into the control system.

## 4. Improvements on resolution and repeatability

Intensity measurements are used to set limit for alarms and to track efficiency. This allows the user to keep the beam within safety and operational requirements. This demands that intensity measurements need to be accurate and repeatable. This was done by identifying noise sources, temperature sensitivities, and timing sensitivities. Consequently, modifications to several aspects of the system were made to all installations through the site.

### 4.1 System hardware modifications

The figure below provides a simplified view of the complete system.

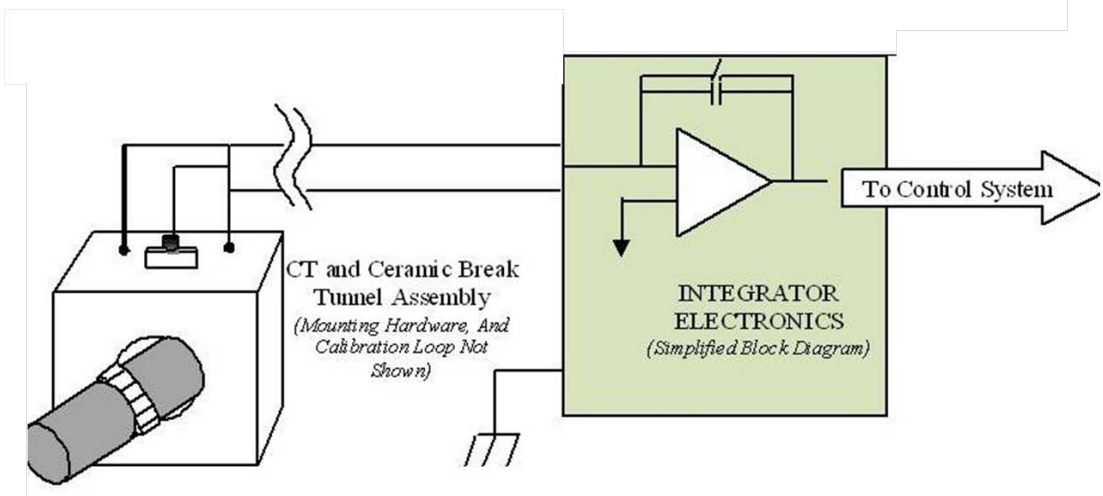

**Figure 1.** Simplified system block diagram (not to scale)

### 4.1.1 Mounting hardware

Previously, a path for low frequency components of the beam image current was provided using conductive braids strapped over the CT. This was replaced with a standardized mounting stand which is measured to fit exactly into a surveyed location. The entire assembly can be supported from either the floor or ceiling (Figure 2).

Conductive side walls of the stand are pressure fitted to clasp around the beam pipe, enclosing both the CT and ceramic break. Depending on the location, either the top or bottom plate completes the electrical path for the low frequency component of the beam image currents. The sides are held tight with screws, and the resistance across each joint is measured. There must be low resistance and good electrical contact between the side walls of the mounting stand and the beam pipe. Typically a good connection is measured to be less than 50μΩ.

The opposing plate is non-conductive, typically made from G10 material. This ensures the CT, any filters or preamplifiers, and any cables are isolated from the beam pipe with minimum capacitance.

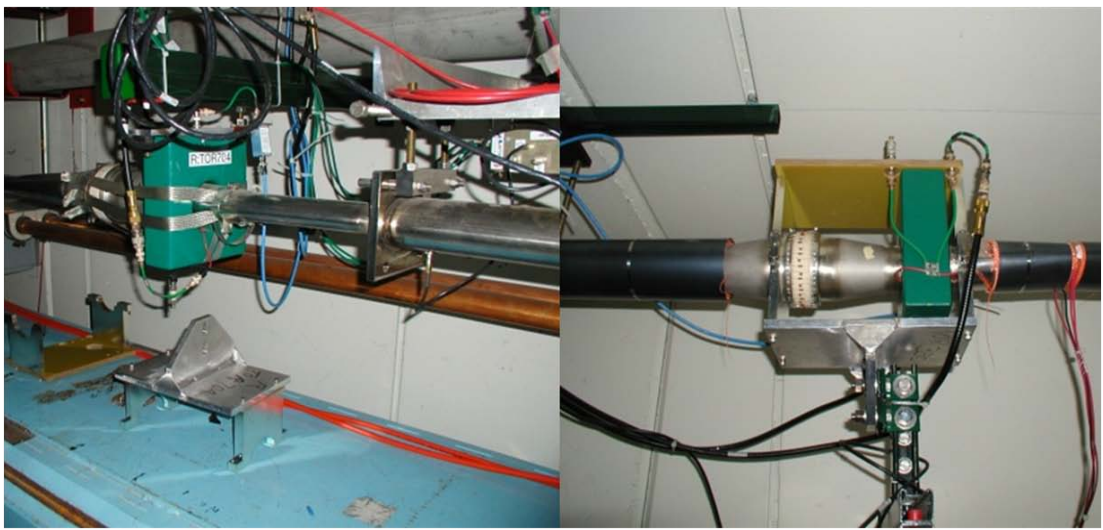

**Figure 2.** Before (left) and after (right) improvements to mounting stands.

### 4.1.2 Gap network

The ceramic break, chosen by the beam pipe diameter, provides a break in the otherwise continuous beam pipe. A passive network is placed across the ceramic break to control the gap impedance as well as minimize any high harmonics frequency leakage from the beam.

Pairs of resistor and capacitor in series are evenly distributed in parallel around the ceramic gap. These pairs were originally installed by soldering the RC pairs to two steel cable ties, which were fastened to each side of the gap. As opportunities arise, these are replaced with a double-sided Kapton flex circuit board with the same passive network in surface mount components. Adjustable hose-clamps pressure fit the board across the gap.

### 4.1.3 Analog conditioning of CT output

The output of the CT is usually conditioned directly at the CT's head. However, if the tunnel location is exposed to high radiation levels, any analog conditioning components are moved outside the tunnel enclosure.

A preamplifier is used to provide an active gain stage to accommodate the electronics input range and the CT's sensitivity. In addition, a passive coaxial inline filter limits the signal bandwidth to frequencies where the CT and active components behave linearly.

The CT output is carried into the tunnel enclosure by low-loss coaxial cables. The length of the cable runs are determined by the placement of the CT with respect to the penetrations. If necessary, a set of ferrite cores can be heat-shrunk around the receiving end of each cable to reduce common-mode noise due to unwanted capacitive coupling.

### 4.1.4 CT Signal Processing

Although there are plans to instrument all future upgrades and installations within a VME electronics environment, most installations use NIM-powered electronics to process the CT's signals. These electronics consist of a switch-capacitor integrator circuit, field programmable gate array (FPGA), crystal oscillator, analog-to-digital converter (ADC), and digital-to-analog converter (DAC).

The integrator circuit is gated by the FPGA using an external TTL pulse that is synchronous with a transfer event. During beam commission, the delay for trigger is adjusted so that the beam pulse arrives within the integration gate. The integrated value is sampled by the ADC and stored in the FPGA as a 16-bit value which is directly proportional to the beam intensity. The full-scale intensity range can be adjusted with the ratio of the input resistor and the feedback capacitor in the integrator circuit. Unless requested by the user, the value is held until the next trigger is received.

Two intensity outputs are provided to the control system, where are scaled into the appropriate units. One output is the 16-bit digital intensity reading, which is typically read by either a VME or a CAMAC digital I/O card. The CAMAC digital I/O card is limited to a 15 Hz maximum update rate. The other output is an analog +/-10V output, typically used by the Controls multiplexed ADC.

#### 4.1.4.1 Electronic hardware modifications

The switch-capacitor integrator circuit was redesigned with several updated components.

Low drop-out power supply regulators were used to minimize power dissipation. Also, both active and passive filters were added to limit both the input signal bandwidth and common-mode bandwidth. Operational amplifiers with better common-mode rejection, low noise figures, and low offset drift were selected. Resistor values were reduced when possible to minimize thermal noise contributions. The selected integrating, feedback capacitors were composed of a polystyrene material; they are characterized with high stability and high insulation resistance, which is ideal when used in integration circuits. Furthermore, the switches used were replaced to reduce charge injection. For a $1\times10^{13}$ full-scale intensity system with an 11 µs integration gate, the RMS noise contribution of the analog section is less than $3\times10^8$.

The main noise sources are the ADC and DAC. The ADC is a capacitor-based sample-and-hold circuit with an 84 dB signal to noise ratio (SNR) and a maximum sampling rate of 250 kHz. The 16-bit DAC, with 83dB signal-to-noise and distortion ratio (SINAD), is fabricated with CMOS logic functions and high precision bipolar linear circuitry. The ADC and DAC each contribute about $2\times10^9$ errors for a $1 \times10^{13}$ full-scale intensity system.

The combination of all electronics modifications improved reduced the RMS noise well as temperature sensitivity of the system. The total RMS noise of the electronics improved by about a factor of three (Figure 3).

In addition, the temperature sensitivity of the electronics was measured to be ±0.01% of full-scale per degree Fahrenheit (Figure 4). The inherent temperature sensitivity of the toroid core is typically ±0.02% error /°F.

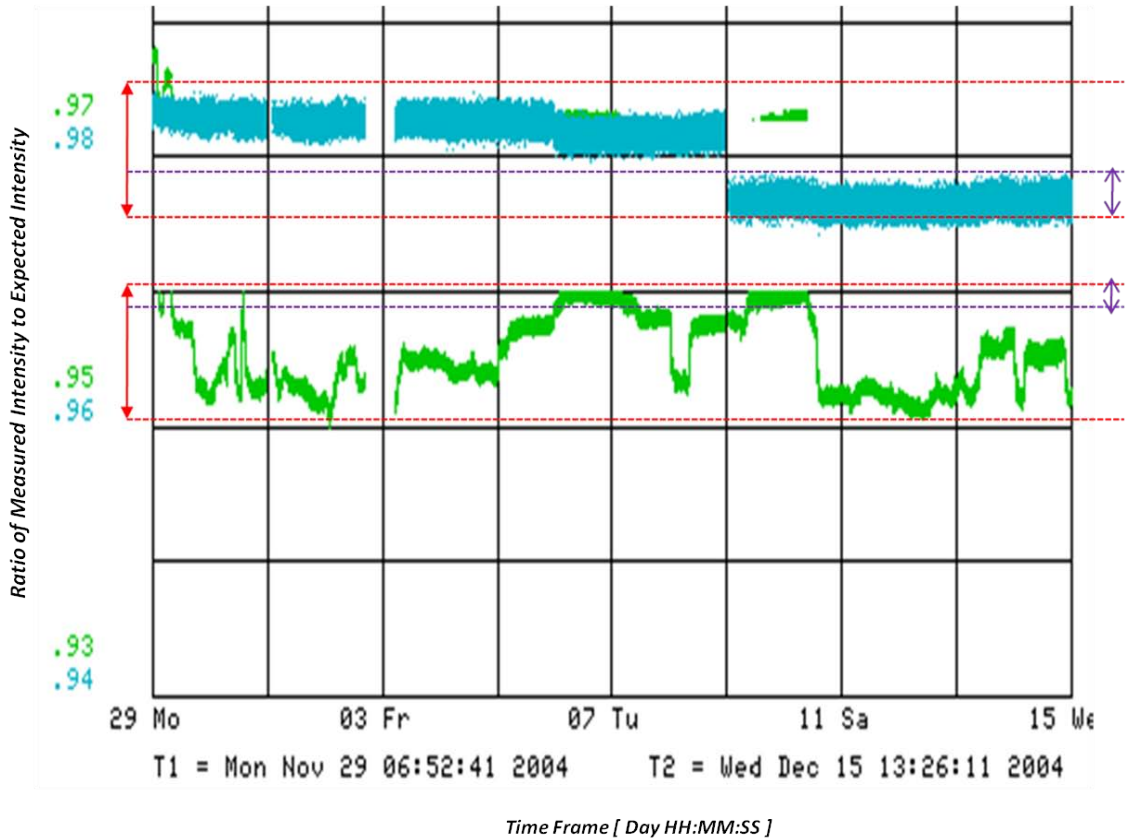

**Figure 3.** The plot above compares fluctuations in gain over time between transport beam intensity monitors with (blue) and without (green) electronic modifications. The gains were tracked during successive calibrations in between beam transfers over several days and were calculated using a least squares fit model between measured intensities and expected intensities. Although both show ±0.5% long-term gain drift (red arrow bars), the new electronics show ±0.1% RMS pulse-to-pulse variation, in contrast to ±0.3% in the older electronics (purple arrow bars).

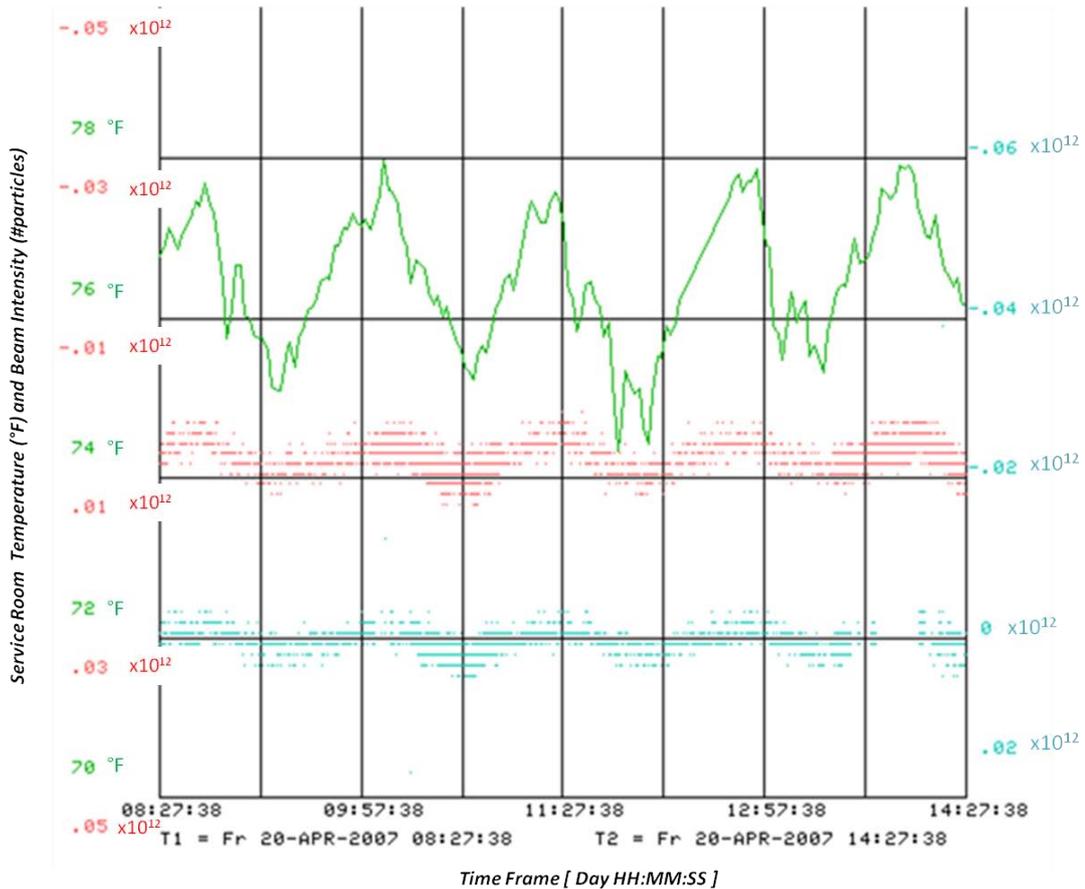

**Figure 4.** Plot showing the temperature correlation between two transport beam intensity monitors (red and cyan) and the room temperature (green). Both toroid monitors are $1 \times 10^{13}$ at full-scale intensity with an 11 μsec integration gate. The plot shows about $3.4 \times 10^9$ variations due to the temperature change.

**4.2 System software modifications**

An FPGA is used to generate the control signal to open or close the switches in the integrator circuit. A gating window for the switched-capacitor circuit is created when a TTL external trigger is received.

The figure below provides a basic timing diagram for the switched capacitor circuit.

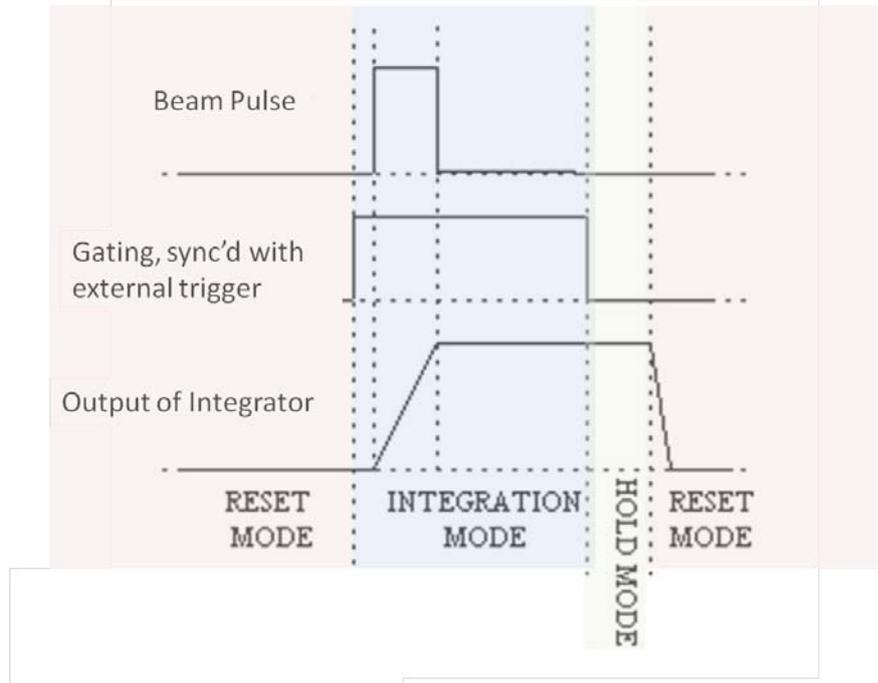

**Figure 5.** Timing diagram for switch capacitor circuit. During the reset mode, the output of the integrator is effectively grounded, providing a zero-state for the system. During the integration mode, any signal "captured" within the gating window is integrated. After the integration window closes, the integrated value is held long enough for an onboard ADC can sample it. This value is stored in an FPGA, which is later available to the control system.

**4.2.1 Gate width selection**

The switched-capacitor circuit is gated with the FPGA to integrate any signal within an integration window. Proper selection of the gate width helped minimize the noise of the system.

The noise output of the integrator circuit is proportional to the square root of the gate width. In addition, spurious noise pickup is proportional to the gate width. So, minimizing the desired gate width for the intended transfer beam pulse width is ideal. Care must be taken, however, to insure that the integration gate fully accommodates the entire beam pulse. In particular, any overshoot in the rising edge of the pulse will be accompanied by a symmetrical tail after the falling edge. Both must be integrated to provide an accurate result. Increasing the time constant relative to a desired minimum gate width can additionally reduced the final noise output. For 1.6 μs beam pulse, the gate width is usually 11 μs wide with an integrating time constant of about 330 ns.

**4.2.2 Timing sensitivity**

The triggering mechanism of the toroid system is synchronized with a beam transfer event, so that the transfer beam pulse will arrive within the integration gate. With a 10 μs integration gate, the maximum external trigger rate is 66.7 kHz.

During the commissioning of the transport line, the trigger delay value is set to allow the beam pulse arrives early in the integration gate. This provides adequate time for any cable dispersion. The sensitivity to timing is typically about 0.14% error per μsec, with 152 meters of cable or less. However, it can increase to as much as 1.3%/μs for longer cable lengths of 366 meters.

### 4.2.3 Baseline-correction algorithm

The onboard FPGA is used to control the gating for the switched-capacitor circuits as well as the ADC and DAC control signals. The FPGA primarily provides a temporary storage location for the integrated value. However, this function was extended to correct for any baseline errors from low frequency common-mode noise signals.

For each external trigger received, two gates of the same width are produced by the FPGA. External TTL triggering controls allow the 2nd gate to capture the pre-pulse baseline, and the first gate, the transferred beam pulse. This gating scheme was chosen to preserve the timing delays previously set during commissioning of the transport lines years ago.

For each gating pair, the FPGA subtracts the integrated "baseline" value from the integrated "sample" value. For each trigger event the integrator receives, it opens a 10 μs gate to sample the transfer, waits ~5 μs, and opens a second gate to sample the baseline in proximity to the beam. The gate separation allows for adequate reset time for the switched-capacitors and ADC acquisition of the integrated baseline value. The integrated sample values are held for ~3.3 μs for ADC acquisition. Consequently, the maximum trigger rate of the electronics with the algorithm is reduced to ~32 kHz.

The addition of the baseline-algorithm improved fluctuations in offsets during no beam events (Figure 6).

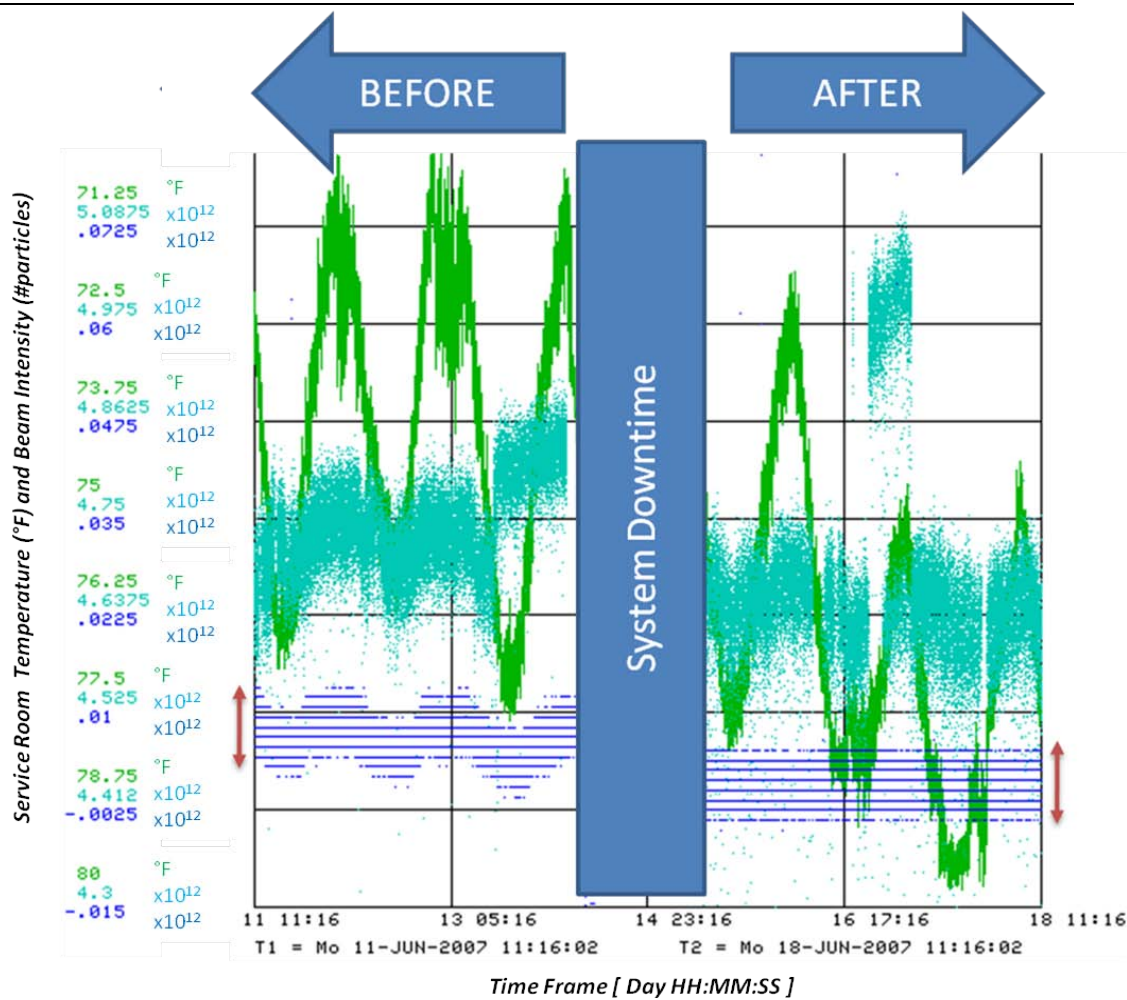

**Figure 6.** Plot showing the effects before (left) and after (right) the implementation of the baseline-correction algorithm. The green trace is a plot of the service room temperature (in degrees Fahrenheit), acting as the noise source. The cyan and blue traces plots the output of a $1 \times 10^{13}$ full scale intensity system during beam transfers, and during no beam events in between transfers, respectively. As seen in the blue trace, the addition of the baseline-correction greatly reduced the impact of the room temperature (red arrow bars).

#### 4.2.3.1  Baseline rejection response

As the frequency of the integrator input signal increases, the baseline rejection decreases linearly. Maximum rejection of about 18dB occurs at 1 kHz. For frequencies >10 kHz, the algorithm becomes a disadvantage. Theoretically, the worst case, in which the baseline is doubled, occurs at ~32 kHz. Fortunately, the maximum sampling rate of the ADC places a zero at ~32 kHz.

#### 4.2.3.2  Sensitivity to trigger rate

The integrator external trigger rate was varied while holding the input constant. With no baseline subtraction algorithm, the intensity measurement shows sensitivity to triggering rates >

1 kHz. The intensity measurement exhibited as much as 44% error of full scale. With the baseline subtraction algorithm, this error was reduced to ±1% error.

## 5. Conclusion

Intensity monitors are common diagnostics tools used by operators to measurement beam intensity in transport lines. During the Run II era, much effort was devoted to improving the system's resolution and repeatability. Observations made at individual installations drove site-wide hardware and software modifications to all the transport line intensity monitors.

## Acknowledgments

Contributions from the following are gratefully acknowledged: R. Webber, J. Crisp, D. Heikkinen, and many others at Pearson Electronics as well as FNAL Accelerator Division's Instrumentation Department, Controls Department, and Operations.

## References

[1] Talman, R., "*Beam Current Monitors*," in the proceedings of the *1st Annual Accelerator Instrumentation Workshop*, 1989, Upton, New York), AIP Conference Proceedings No. 212, ISBN O-883 18-645-4, 1990, pp. l-25.

[2] Robert C. Webber, *Charged Particle Beam Current Monitoring Tutorial ,* in proceedings of $6^{th}$ *Beam Instrumentaiton Workshop (BIW 1994), October 2 -6, 1994* Vancouver, Canada

[3] Robert C. Webber, *A Tutorial on beam current monitoring,* in proceedings of $9^{th}$ *Beam Instrumentaiton Workshop (BIW 2000), May 8-11,* 2000 Cambridge, MA

[4] Robert C. Webber, *A Tutorial on non-intercepting electromagnetic monitors for charged particle beams,* in proceedings of $8^{th}$ *International Meeting on Nuclear Applications and Utilization of Accelerators(AccAPP'07)*, 30 July – 2 August, 2007 Pocatello, ID.

[5] http://www.pearsonelectronics.com/